%% file: main.tex
\documentclass[10pt,twocolumn,letterpaper]{article}

\usepackage[pagenumbers]{cvpr}      
\usepackage[table]{xcolor}
\usepackage{booktabs}
\usepackage{graphicx}
\usepackage[accsupp]{axessibility}  
\input{preamble}

\definecolor{cvprblue}{rgb}{0.21,0.49,0.74}
\usepackage[pagebackref,breaklinks,colorlinks,allcolors=cvprblue]{hyperref}

\def\paperID{*****} 
\def\confName{CVPR}
\def\confYear{2026}

\title{BRIDGE: Multimodal-to-Text Retrieval via Reinforcement-Learned Query Alignment}

\author{
Mohamed Darwish Mounis$^{1}$ \qquad
Mohamed Mahmoud$^{2,3}$ \qquad 
Shaimaa Sedek$^{3}$ \qquad  \\
Mahmoud Abdalla$^{2}$ \qquad
Mahmoud SalahEldin Kasem$^{2,3}$ \qquad
Abdelrahman Abdallah$^{3,4}$ \qquad  \\
Hyun-Soo Kang$^{1}$\thanks{Corresponding author.}
\\[0.5em]
$^{1}$ High institute for computer \& information systems
$^{2}$ Chungbuk National University \\
$^{3}$ Assiut University
$^{4}$ University of Innsbruck
\\[0.5em]
}

\begin{document}
\maketitle
\input{sec/0_abstract}    
\input{sec/1_intro}
\input{sec/2_relatedwork}

\input{sec/3_model}
\input{sec/4_Experiments}
\input{sec/5-result}

\input{sec/6_Conclusion}

{
    \small
    \bibliographystyle{ieeenat_fullname}
    \bibliography{main}
}


\end{document}

%% file: preamble.tex
\usepackage{booktabs}
\usepackage{multirow}
\usepackage{array}



%% file: sec/0_abstract.tex
\begin{abstract}
Multimodal retrieval systems struggle to resolve image-text queries against 
text-only corpora: the best vision-language encoder achieves only 27.6 nDCG@10 
on MM-BRIGHT, underperforming strong text-only retrievers. We argue the 
bottleneck is not the retriever but the query --- raw multimodal queries 
entangle visual descriptions, conversational noise, and retrieval intent in 
ways that systematically degrade embedding similarity. We present \textbf{BRIDGE}, a two-component system that resolves this mismatch 
without multimodal encoders. \textbf{FORGE} (\textbf{F}ocused Retrieval Query 
Generato\textbf{r}) is a query alignment model trained 
via reinforcement learning, which distills noisy multimodal queries into 
compact, retrieval-optimized search strings. \textbf{LENS} 
(\textbf{L}anguage-\textbf{E}nhanced \textbf{N}eural \textbf{S}earch) is a 
reasoning-enhanced dense retriever fine-tuned on reasoning-intensive retrieval 
data to handle the intent-rich queries FORGE produces. Evaluated on MM-BRIGHT (2,803 queries, 29 domains), BRIDGE achieves 
\textbf{29.7} nDCG@10, surpassing all multimodal encoder baselines including 
Nomic-Vision (27.6). When FORGE is applied as a plug-and-play aligner on top 
of Nomic-Vision, the combined system reaches \textbf{33.3} nDCG@10 --- 
exceeding the best text-only retriever (32.2) --- demonstrating that 
\textit{query alignment} is the key bottleneck in multimodal-to-text 
retrieval. \footnote{\url{https://github.com/mm-bright/multimodal-reasoning-retrieval}}
\end{abstract}

%% file: sec/1_intro.tex
\section{Introduction}

The ability to retrieve relevant information from large text corpora is 
fundamental to knowledge-intensive applications such as question answering, 
retrieval-augmented generation, and agentic systems~\cite{lewis2020rag}.  
Recent advances in dense retrieval have yielded powerful embedding models 
that capture rich semantic relationships between queries and 
documents~\cite{karpukhin2020dpr,abdalla2025think,abdallah2025dear,ali2026recor,ni2022large,abdallah2024arabicaqa}, and in visual modeling, such as face unmasking~\cite{mahmoud2026m2unet}, improve visual representation but do not address retrieval over large text corpora.
 However, real-world 
queries are increasingly multimodal. Users post screenshots of error 
messages, attach circuit diagrams, or include charts from technical reports 
when seeking help online. In these settings, retrieval requires not just 
semantic matching but \textit{reasoning} --- understanding what an image 
depicts, how it relates to the text question, and which documents in the 
corpus address both dimensions jointly. This challenge is starkly reflected in MM-BRIGHT~\cite{abdallah2026mmbright}, 
the first benchmark for reasoning-intensive multimodal retrieval. Despite significant advances in vision-language models, the best multimodal 
retriever achieves only 27.6 nDCG@10 --- \textit{lower} than the best 
text-only retriever (32.2). This gap is not merely counterintuitive; it 
reveals a fundamental problem: multimodal queries are \textit{noisy}. 
Crucially, text-only retrievers succeed precisely because they receive 
\textit{clean, intent-focused} queries --- the very property that 
multimodal queries lack. BRIDGE is designed to close this gap by 
transforming noisy multimodal queries into the kind of clean text queries 
that strong retrievers already handle well. They entangle image captions, 
conversational context, task-irrelevant background, and retrieval intent 
into a single unstructured input that confuses dense embedding models. The 
retriever does not fail because it lacks visual understanding; it fails 
because the query it receives is a poor representation of what the user 
actually needs to find. 

\begin{figure*}[t]
    \centering
    \includegraphics[width=0.9\linewidth]{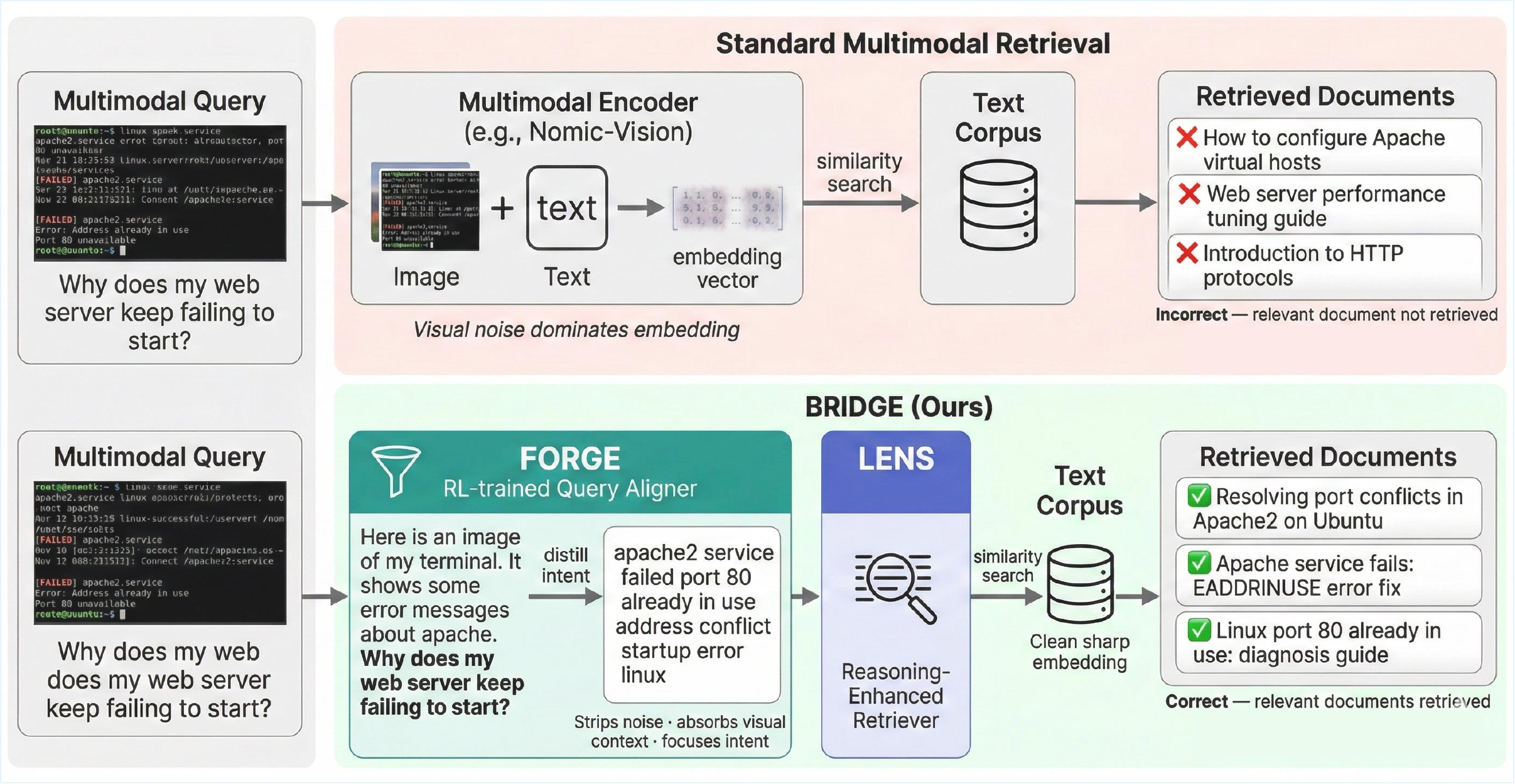}
    \caption{An example of BRIDGE in action. The raw multimodal query 
    mixes conversational context and visual content, producing a noisy 
    embedding that fails to retrieve the correct document. FORGE distills 
    the query into a compact, intent-focused search string that LENS 
    successfully resolves against the text corpus.}
    \label{fig:bridge_example}
\end{figure*}

Existing approaches to this problem focus on the retriever side: training 
larger vision-language encoders~\cite{faysse2024colpali, zhang2024gme, 
jiang2024vlm2vec}, fine-tuning on multimodal contrastive 
data~\cite{zhai2023siglip}, or applying LLM-based 
reranking~\cite{sun2023rankgpt,abdallah2025rankify,mozafari2025good,abdallah2025dear}. However, all of these approaches accept 
the noisy query as a fixed input and attempt to overcome it through model 
capacity. None address the root cause: the query itself must be restructured 
before retrieval. As illustrated in Figure~\ref{fig:bridge_example}, when a 
user query combines an image of a terminal error log with the text 
``\textit{why does my service keep failing?}'', no amount of visual encoding 
can compensate for the fact that the query embedding is dominated by 
conversational noise rather than the specific error type visible in the image.

We introduce \textbf{BRIDGE}, a two-component system that addresses the 
multimodal retrieval gap at its source: the query. The first component, 
\textbf{FORGE} (\textbf{F}ocused Retrieval Query Generato\textbf{r}), is 
a query alignment model fine-tuned from Qwen2.5-7B-Instruct via 
reinforcement learning on curated multimodal retrieval data. Unlike 
heuristic query rewriting approaches, FORGE is trained to optimize for 
downstream retrieval quality: given a raw multimodal query comprising a 
text question and an image caption, it learns to strip linguistic noise, 
absorb the visual context, and produce a compact, dense-retrieval-optimized 
search string that faithfully represents the user's core intent. The second 
component, \textbf{LENS} (\textbf{L}anguage-\textbf{E}nhanced \textbf{N}eural 
\textbf{S}earch), is a reasoning-enhanced dense retriever fine-tuned on 
reasoning-intensive retrieval data to handle the intent-rich, structured 
queries that FORGE produces. Together, BRIDGE requires no multimodal encoder 
at inference time --- it operates entirely in the text modality, making it 
lightweight, modular, and scalable.

BRIDGE is motivated by a simple insight: \textit{the modality gap in 
retrieval is primarily a query representation problem, not a model capacity 
problem.} FORGE learns to bridge what the user expresses multimodally and 
what the retriever needs as input; LENS is trained to retrieve effectively 
given that bridge.

Our contributions are as follows:
\begin{itemize}
    \item We identify that the core bottleneck in multimodal-to-text retrieval 
    is \textit{query representation noise} --- the entanglement of visual 
    descriptions, conversational context, and retrieval intent --- rather 
    than retriever model capacity.

    \item We introduce \textbf{FORGE}, a reinforcement-learned query alignment 
    model trained on Qwen2.5-7B-Instruct that distills noisy multimodal queries 
    into compact, retrieval-optimized search strings, eliminating the need for 
    multimodal encoders at inference time.

    \item We introduce \textbf{LENS}, a reasoning-enhanced dense retriever 
    fine-tuned on reasoning-intensive retrieval data to handle the 
    intent-rich queries produced by FORGE.

    \item We demonstrate that \textbf{BRIDGE} (FORGE + LENS) achieves 
    \textbf{29.7} nDCG@10 on MM-BRIGHT's multimodal-to-text track, 
    surpassing all multimodal encoder baselines across all 29 domains. 
    As a plug-and-play aligner, FORGE further boosts Nomic-Vision from 
    27.6 to \textbf{33.3} --- surpassing the best text-only retriever 
    (32.2) --- demonstrating that query alignment is the missing 
    ingredient in multimodal retrieval, regardless of the retriever used.
\end{itemize}

%% file: sec/2_relatedwork.tex
\section{Related Work}

\subsection{Dense Retrieval and Reasoning-Intensive Benchmarks}

Dense retrieval using bi-encoder models has become the dominant paradigm for 
large-scale information retrieval~\cite{karpukhin2020dpr, reimers2019sentence}. 
These models independently encode queries and documents into a shared embedding 
space and retrieve via efficient nearest-neighbor search~\cite{johnson2019faiss}. 
While highly effective on fact-seeking benchmarks such as BEIR~\cite{thakur2021beir} 
and MTEB~\cite{muennighoff2023mteb}, bi-encoders degrade sharply when relevance 
requires multi-step reasoning rather than surface-level semantic matching. 
BRIGHT~\cite{su2025bright} RECOR~\cite{ali2026recor} and TEMPO~\cite{abdallah2026tempo} exposed this 
limitation clearly: even the strongest embedding models collapse from 59.0 nDCG@10 
on BEIR to 18.3 on reasoning-intensive queries. Subsequent work has sought to 
address this gap through reasoning-aware fine-tuning~\cite{shao2025reasonir, 
das2025rader} and iterative query expansion~\cite{wang2023query2doc, lei2025thinkqe}. 
%

\subsection{Multimodal Embedding Models}

Contrastive vision-language models such as CLIP~\cite{radford2021clip} and 
SigLIP~\cite{zhai2023siglip} established the foundation for joint image-text 
embedding through large-scale contrastive pretraining. More recent work has 
leveraged the representational power of Multimodal LLMs for retrieval: 
VLM2Vec~\cite{jiang2024vlm2vec} demonstrated that instruction-tuned VLMs can 
be converted into powerful embedding models through contrastive training on the 
MMEB benchmark, while GME~\cite{zhang2024gme} extended this to support 
any-to-any retrieval across text, image, and fused modalities. Nomic Embed 
Vision~\cite{nomic2024vision} shares an embedding space between a vision 
encoder and a text model, enabling zero-shot multimodal retrieval. Despite 
these advances, all embedding-based models share a fundamental limitation: 
they cannot reason about what a query image \textit{implies} for document 
relevance. MM-BRIGHT~\cite{abdallah2026mmbright} confirmed this directly --- 
the best multimodal model (27.6 nDCG@10) underperforms the best text-only 
retriever (32.2), revealing that the bottleneck is not visual encoding capacity 
but query representation quality. 

\subsection{Visual Document Retrieval}

ColPali~\cite{faysse2024colpali}, DSE~\cite{ma2024dse}, and 
VisRAG~\cite{yu2024visrag} embed full document pages as images for 
dense retrieval. This line of work assumes documents are visual, operating 
in a fundamentally different setting from ours where the corpus is 
text-only and only the query contains visual content.

\subsection{Query Rewriting and Alignment}

Query rewriting has a long history in information retrieval, from classical 
pseudo-relevance feedback~\cite{robertson2009bm25} to modern LLM-based 
reformulation. Ma et al.~\cite{ma2023rewrite} introduced the 
\textit{Rewrite-Retrieve-Read} framework, proposing that adapting the query 
itself --- rather than the retriever or reader --- can close the gap between 
user intent and retrieval performance. They trained a small rewriter using 
reinforcement learning feedback from a downstream reader, establishing the 
paradigm of training-based query alignment. HyDE~\cite{gao2022hyde} generates 
hypothetical documents from the query for zero-shot dense retrieval, while 
Query2Doc~\cite{wang2023query2doc} expands queries with pseudo-documents 
through few-shot prompting. Re-Invoke~\cite{chen2024reinvoke} applies LLMs 
to extract underlying user intents from verbose queries before retrieval, 
with a particular focus on tool retrieval settings.

In the conversational search domain, query rewriting has been used to resolve 
coreferences and compress context into standalone queries~\cite{yu2020conqrr, 
mao2023convgqr}. More recent work has studied rewriting specifically for 
reasoning-intensive retrieval: ThinkQE~\cite{lei2025thinkqe} generates 
chain-of-thought expanded queries, while DIVER~\cite{long2025diver} integrates 
iterative query expansion with document feedback and hybrid reranking into a 
unified pipeline.

\subsection{Reinforcement Learning for Query Optimization}

Reinforcement learning has recently emerged as a powerful paradigm for query 
optimization, motivated by the success of RLHF in LLM 
alignment~\cite{ouyang2022rlhf} and reasoning~\cite{deepseek2025r1}. The core 
insight is that query quality is best measured by retrieval outcome --- a 
signal that is available without human annotation and directly optimizable via 
policy gradient methods.

DeepRetrieval~\cite{jiang2025deepretrieval} pioneered RL-based query 
generation using retrieval recall as reward without supervised references. 
RL-QR~\cite{cha2025rlqr} demonstrated up to $3.9{\times}$ gains on lexical 
retrievers via verifiable search rewards. ConvSearch-R1~\cite{convr1} applied 
GRPO to conversational reformulation without human rewrites. Ma et 
al.~\cite{ma2023rewrite} trained a rewriter with RL feedback from a reader, 
while Nguyen et al.~\cite{nguyen2025rlecom} applied RL-based rewriting in 
e-commerce via simulated user feedback. FORGE extends this paradigm to the 
multimodal setting, uniquely optimizing for vision-grounded query distillation 
rather than text-only expansion.

%% file: sec/3_model.tex
\section{Method}

\subsection{Problem Formulation}

We address the \textit{multimodal-to-text} retrieval task. Let 
$\mathcal{C} = \{d_1, d_2, \ldots, d_N\}$ denote a corpus of $N$ text-only 
documents. A multimodal query is a pair $q = (q_t, q_v)$, where $q_t \in 
\mathcal{V}^*$ is a natural language question and $q_v$ is an associated 
image (e.g., a diagram, screenshot, or chart). The objective is to retrieve 
a ranked list of $k$ documents $\hat{\mathcal{D}}_k \subset \mathcal{C}$ 
that are most relevant to the full intent expressed by $(q_t, q_v)$.

Standard dense retrievers compute relevance as:
\begin{equation}
    \text{score}(q, d_i) = \cos\bigl(\phi(q_t, q_v),\ \psi(d_i)\bigr)
\end{equation}
where $\phi(\cdot)$ and $\psi(\cdot)$ are query and document encoders.This formulation fails in the multimodal-to-text setting because the 
encoder $\phi$ captures the \textit{surface form} of the noisy multimodal 
input rather than the latent retrieval intent. We address this not by 
improving $\phi$, but by transforming $(q_t, q_v)$ into a clean, 
retrieval-optimized text query \textit{before} any embedding is computed.

\subsection{BRIDGE Overview}

\begin{figure*}[t]
    \centering
    \includegraphics[width=0.8\textwidth]{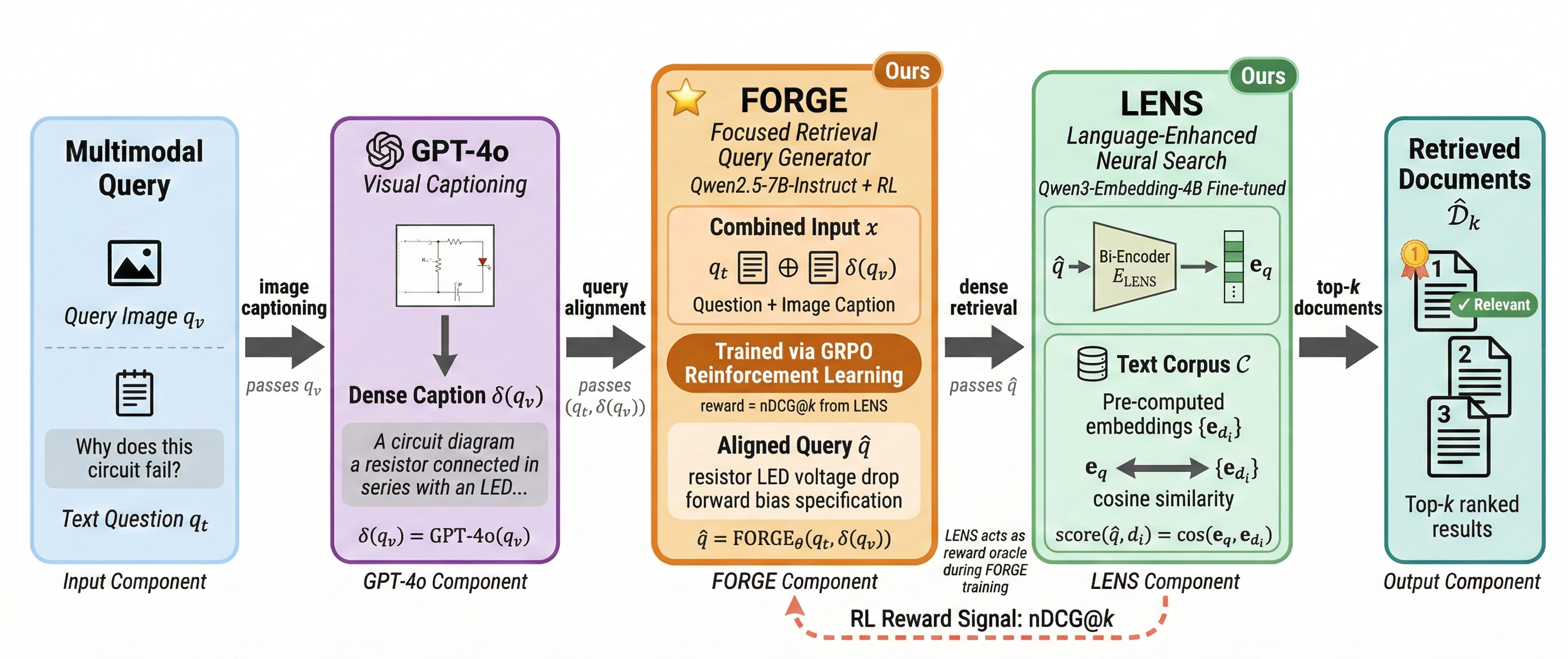}
    \caption{Overview of the BRIDGE framework. Given a multimodal query 
    $(q_t, q_v)$, GPT-4o first generates a dense image caption $\delta(q_v)$. 
    FORGE then receives the concatenated input $(q_t, \delta(q_v))$ and 
    distills it into a compact, retrieval-optimized query $\hat{q}$ via 
    reinforcement-learned alignment. LENS encodes $\hat{q}$ and retrieves 
    the final ranked documents from the text corpus. No multimodal encoder 
    is used at retrieval time.}
    \label{fig:bridge_pipeline}
\end{figure*}

BRIDGE resolves multimodal queries against text corpora through a three-stage 
pipeline, illustrated in Figure~\ref{fig:bridge_pipeline}:

\begin{enumerate}
    \item \textbf{Visual Captioning (GPT-4o).} The query image $q_v$ is 
    converted into a dense textual description $\delta(q_v)$ using GPT-4o, 
    grounding the visual content in natural language.
    
    \item \textbf{Query Alignment (FORGE).} The raw query pair $(q_t, 
    \delta(q_v))$ is fed into FORGE, a reinforcement-learned query alignment 
    model that produces a compact, retrieval-optimized search string 
    $\hat{q} = \text{FORGE}(q_t, \delta(q_v))$.
    
    \item \textbf{Retrieval (LENS).} The aligned query $\hat{q}$ is encoded 
    by LENS, a reasoning-enhanced dense retriever, to retrieve the final 
    ranked list $\hat{\mathcal{D}}_k$ from $\mathcal{C}$.
\end{enumerate}

The full pipeline can be expressed as:
\begin{equation}
    \hat{\mathcal{D}}_k = \text{LENS}\Bigl(\text{FORGE}\bigl(q_t,\ 
    \underbrace{\text{GPT-4o}(q_v)}_{\delta(q_v)}\bigr),\ \mathcal{C},\ k\Bigr)
\end{equation}

Crucially, once FORGE produces $\hat{q}$, the entire retrieval process 
operates in the text modality. No visual encoder is invoked at retrieval 
time, making BRIDGE lightweight, modular, and compatible with any text 
retriever.

\subsection{ Visual Captioning}
\label{sec:captioning}

The query image $q_v$ encodes domain-specific visual information --- 
molecular structures, UI screenshots, circuit diagrams, data charts --- that 
is inaccessible to text-only models. We convert $q_v$ into a dense textual 
description using GPT-4o:
\begin{equation}
    \delta(q_v) = \text{GPT-4o}\bigl(\textsc{CaptionPrompt}(q_v)\bigr)
\end{equation}

The \textsc{CaptionPrompt} instructs GPT-4o to produce a comprehensive, 
domain-aware description capturing object types, spatial relationships, 
labels, and structural features. The full multimodal query context is 
then represented as:
\begin{equation}
    x = \bigl[q_t\ \|\ \texttt{Image Description: }\ \delta(q_v)\bigr]
\end{equation}
where $\|$ denotes string concatenation, and $x$ is the input to FORGE.

\subsection{ Focused Retrieval Query Generator}
\label{sec:forge}

The combined input $x$ is a faithful but unstructured representation of the 
user's intent: it contains conversational preamble, visual details, and 
domain context mixed together. Dense retrievers are sensitive to this 
entanglement --- they embed the query as a whole, and lexical noise dominates 
the resulting vector. FORGE is trained to solve this: given $x$, it produces 
a compact, distilled query $\hat{q}$ that retains only the retrieval-critical 
information.

FORGE is built on Qwen2.5-7B-Instruct~\cite{qwen2025}, an instruction-tuned 
language model. The model takes $x$ as input under a system prompt that 
defines the task as query alignment for dense retrieval:
\begin{equation}
    \hat{q} = \text{FORGE}_\theta(x) = \text{FORGE}_\theta(q_t, \delta(q_v))
\end{equation}
where $\theta$ denotes the model parameters. The output $\hat{q}$ is 
constrained to be a concise search string of no more than 200 words, 
containing only the essential keywords and semantic concepts needed for 
dense retrieval. To train FORGE, we construct a curated dataset of multimodal retrieval 
pairs. Each training instance consists of a multimodal query $(q_t, q_v)$, 
its GPT-4o-generated caption $\delta(q_v)$, and a ground-truth relevant 
document $d^+ \in \mathcal{C}$. 
For each instance, we also sample hard negative documents 
$d^- \in \mathcal{C} \setminus \{d^+\}$ using BM25 and LENS initial 
retrieval to provide informative contrastive reward signals during RL training.

Unlike supervised approaches that train FORGE to mimic reference rewrites, 
we train directly for retrieval quality using RL with a retrieval-based 
reward. The policy is the FORGE model $\pi_\theta$, the action is the 
generated query $\hat{q}$, and the reward $r(\hat{q}, d^+)$ is defined as:
\begin{equation}
    r(\hat{q}, d^+) = \text{nDCG@}k\bigl(\text{LENS}(\hat{q}, \mathcal{C}),\ 
    \{d^+\}\bigr)
\end{equation}

That is, FORGE is rewarded for generating queries that cause LENS to rank 
the ground-truth document highly. We optimize $\pi_\theta$ using Group 
Relative Policy Optimization (GRPO)~\cite{shao2024grpo}:
\begin{equation}
    \mathcal{J}(\theta) = \mathbb{E}_{\hat{q} \sim \pi_\theta(\cdot | x)} 
    \Bigl[r(\hat{q}, d^+) - r_{\text{baseline}}\Bigr] 
    \nabla_\theta \log \pi_\theta(\hat{q} | x)
\end{equation}
where $r_{\text{baseline}}$ is the group-relative baseline computed across 
multiple sampled outputs for the same input $x$. This formulation allows 
FORGE to explore the query rewriting space freely and converge on strategies 
that maximize downstream retrieval performance, without requiring any 
human-annotated reference rewrites.

The RL training loop alternates between:
\begin{enumerate}
    \item \textbf{Rollout}: Sample $G$ candidate queries $\{\hat{q}^{(g)}\}_{g=1}^G$ 
    from $\pi_\theta(\cdot | x)$.
    \item \textbf{Reward}: Compute $r(\hat{q}^{(g)}, d^+)$ for each candidate 
    by running LENS over $\mathcal{C}$.
    \item \textbf{Update}: Apply GRPO gradient update using the group-normalized 
    advantages $A^{(g)} = r(\hat{q}^{(g)}, d^+) - \frac{1}{G}\sum_{g'} 
    r(\hat{q}^{(g')}, d^+)$.
\end{enumerate}

\subsection{ Language-Enhanced Neural Search}
\label{sec:lens}

LENS is a bi-encoder dense retriever built on Qwen3-Embedding-4B. Given an 
aligned query $\hat{q}$, LENS encodes it into a dense vector using the 
hidden state of the \texttt{[EOS]} token from the final layer:
\begin{equation}
    \mathbf{e}_q = E_{\text{LENS}}(\hat{q}) \in \mathbb{R}^d
\end{equation}
Documents are similarly encoded offline:
\begin{equation}
    \mathbf{e}_{d_i} = E_{\text{LENS}}(d_i) \in \mathbb{R}^d, \quad 
    \forall d_i \in \mathcal{C}
\end{equation}
Retrieval is performed via cosine similarity:
\begin{equation}
    \text{score}(\hat{q}, d_i) = \frac{\mathbf{e}_q \cdot \mathbf{e}_{d_i}}
    {\|\mathbf{e}_q\| \cdot \|\mathbf{e}_{d_i}\|}
\end{equation}
and the top-$k$ documents are returned as $\hat{\mathcal{D}}_k = 
\text{argtop-}k_{d_i \in \mathcal{C}} \ \text{score}(\hat{q}, d_i)$.

LENS is fine-tuned on reasoning-intensive retrieval data using contrastive 
learning with in-batch negatives and hard negatives. For each training 
instance $(\hat{q}, d^+, \{d^-_j\}_{j=1}^M)$, we minimize the InfoNCE loss:
\begin{equation}
    \mathcal{L}_{\text{LENS}} = -\log \frac{
        e^{\text{score}(\hat{q}, d^+) / \tau}
    }{
        e^{\text{score}(\hat{q}, d^+) / \tau} + 
        \displaystyle\sum_{j=1}^{M} e^{\text{score}(\hat{q}, d^-_j) / \tau}
    }
\end{equation}
where $\tau$ is the temperature hyperparameter. Hard negatives are mined 
using BM25 and an initial LENS checkpoint to ensure informative contrastive 
signal. The training data spans reasoning-intensive domains including 
mathematics, science, medicine, law, and software engineering to ensure 
robust generalization to the complex queries that FORGE produces.






%% file: sec/4_Experiments.tex
\section{Experiments}

\subsection{Dataset}

We evaluate BRIDGE on MM-BRIGHT~\cite{abdallah2026mmbright}, the first 
reasoning-intensive multimodal retrieval benchmark. MM-BRIGHT consists of 
\textbf{2,803 queries} spanning \textbf{29 technical domains}, including 
Gaming, Chemistry, Law, Sustainability, Earth Science, Mathematics, Computer 
Science, Medicine, and others. Each query is a multimodal pair $(q_t, q_v)$ 
comprising a text question and one or more associated images (diagrams, charts, 
screenshots, molecular structures, etc.), paired with a text-only document corpus.

\subsection{Experimental Setup}

Both FORGE and LENS are trained on \textbf{4$\times$ NVIDIA H100 80GB GPUs} 
using distributed data-parallel training. Visual captioning is performed 
offline using \textbf{GPT-4o} with temperature $= 0$ 
for deterministic and reproducible captions. All captions are generated once 
per query and cached.

\paragraph{FORGE Training.}
FORGE is fine-tuned from \textbf{Qwen2.5-7B-Instruct} using GRPO 
reinforcement learning with $G{=}8$ rollouts per input, learning rate 
$1{\times}10^{-6}$, and a maximum output length of 256 tokens, 
trained for 3 epochs over the MM-BRIGHT training split.

\paragraph{LENS Training.}
LENS is fine-tuned from \textbf{Qwen3-Embedding-4B} using contrastive 
learning with in-batch negatives and $M = 7$ hard negatives per query, 
temperature $\tau = 0.02$, batch size 512, and learning rate 
$1 \times 10^{-5}$ for 3 epochs over reasoning-intensive retrieval data 
spanning mathematics, science, medicine, law, and software engineering.

\subsection{Baselines}

We compare BRIDGE against the following multimodal retrieval baselines, 
all evaluated on the MM-BRIGHT multimodal-to-text track.

\paragraph{Multimodal Retrievers.}
These models encode the full multimodal query $(q_t, q_v)$ into a shared 
embedding space:
\begin{itemize}
    \item \textbf{CLIP}~\cite{radford2021clip}: Contrastive image-text model 
    with a shared embedding space trained on large-scale image-caption pairs.
    \item \textbf{SigLIP}~\cite{zhai2023siglip}: Sigmoid-based contrastive VLM 
    with improved image-text alignment over standard softmax objectives.
    \item \textbf{Jina-CLIP}~\cite{koukounas2024jinaclip}: Multi-task 
    contrastive model supporting both text-image and text-text retrieval.
    \item \textbf{Nomic Embed Vision}~\cite{nomic2024vision}: Shares an 
    embedding space between a vision encoder and a strong text model, enabling 
    zero-shot multimodal retrieval; the strongest multimodal baseline on MM-BRIGHT.
    \item \textbf{BGE-VL}~\cite{bge2024vl}: Multimodal embedding model from 
    the BGE family supporting fused-modal retrieval.
    \item \textbf{GME-Qwen2-VL-2B / 7B}~\cite{zhang2024gme}: Universal 
    multimodal embedding models built on Qwen2-VL at 2B and 7B parameter 
    scales, supporting any-to-any retrieval.
\end{itemize}

\paragraph{Metrics.}
We evaluate retrieval performance using \textbf{Normalized Discounted 
Cumulative Gain at rank 10} (nDCG@10), the 
primary metric of the MM-BRIGHT benchmark. 

%% file: sec/5-result.tex
\section{Results}

\subsection{Main Results.}
Table~\ref{tab:mmbright_results} 
report aggregated and per-domain nDCG@10 across all 29 MM-BRIGHT domains. 
BRIDGE achieves \textbf{29.7} nDCG@10, surpassing the strongest multimodal 
baseline (Nomic-Vision: 27.6) by \textbf{+2.1 points} and outperforming all 
other VLM-based retrievers --- GME-7B (22.0), Jina-CLIP (23.0), SigLIP 
(10.8), CLIP (10.4), and BGE-VL (10.0) --- while operating \textit{entirely 
in the text modality at inference time}. Notably, all contrastive VLM 
baselines score below 28, consistent with the MM-BRIGHT finding that 
embedding-based visual fusion is insufficient for reasoning-intensive 
multimodal retrieval. We note that the best \textit{text-only} retriever on 
MM-BRIGHT reports 32.2 nDCG@10~\cite{abdallah2026mmbright}; however, that 
system operates on text queries only and cannot process image inputs at all. 
BRIDGE uniquely handles multimodal queries containing images while 
approaching this text-only ceiling, reducing the modality gap from 4.6 points 
(Nomic-Vision vs.\ text-only) to 2.5 points --- a \textbf{46\% reduction} 
in the multimodal retrieval gap.

\begin{table*}[t]
\centering
\resizebox{0.8\textwidth}{!}{%
\begin{tabular}{l c c c c c c c c}
\toprule
\textbf{DOMAIN} & \textbf{BGE-VL} & \textbf{CLIP} & \textbf{GME-2B} & 
\textbf{GME-7B} & \textbf{JINA-CLIP} & \textbf{NOMIC} & \textbf{SIGLIP} & 
\textbf{BRIDGE} \\
\midrule
\rowcolor{teal!20} \multicolumn{9}{c}{\textit{STEM \& Life Sciences}} \\
\midrule
Acad   & 4.2  & 4.8  & 16.2 & 27.6 & 22.3 & 22.6 & 3.6  & \textbf{27.7} \\
Bio    & 5.7  & 14.8 & 22.9 & 15.2 & 20.5 & 26.9 & 11.9 & \textbf{41.1} \\
Chem   & 10.8 & 9.6  & 27.2 & 21.9 & 30.6 & 30.6 & 11.6 & \textbf{34.3} \\
Phys   & 6.8  & 6.1  & 13.3 & 14.0 & 14.4 & 17.2 & 7.3  & \textbf{20.7} \\
Math   & 13.1 & 17.9 & 16.4 & 9.3  & 27.0 & \textbf{34.0} & 15.3 & 31.9 \\
Earth  & 10.1 & 10.9 & 20.5 & 26.2 & 24.6 & 30.1 & 11.8 & \textbf{32.3} \\
BioAc  & 13.3 & 11.4 & 10.5 & 13.4 & 19.4 & 23.4 & 14.8 & \textbf{26.4} \\
BioInf & 11.6 & 9.4  & 21.1 & 19.2 & 23.7 & \textbf{33.8} & 16.8 & 30.8 \\
Med    & 12.6 & 9.8  & 22.7 & 19.0 & 26.8 & \textbf{33.9} & 9.1  & 31.3 \\
\midrule
\rowcolor{teal!20} \multicolumn{9}{c}{\textit{Software \& Technical Systems}} \\
\midrule
Apple  & 7.2  & 12.3 & 23.9 & 17.0 & 24.3 & \textbf{28.7} & 4.4  & 21.3 \\
Ubuntu & 11.6 & 5.5  & 25.9 & 34.2 & 26.1 & 34.3 & 12.6 & \textbf{50.4} \\
BTC    & 8.9  & 8.3  & 18.2 & 19.6 & 22.6 & 22.7 & 10.0 & \textbf{28.0} \\
Crypto & 11.3 & 14.8 & 9.8  & 7.1  & 15.5 & 22.4 & 10.2 & \textbf{31.4} \\
QC     & 4.5  & 2.6  & 5.9  & 5.6  & 10.8 & \textbf{12.1} & 2.6  & 8.9  \\
Robot  & 16.1 & 10.6 & 15.8 & 18.7 & 19.0 & \textbf{30.3} & 14.3 & 19.0 \\
Sales  & 14.2 & 2.3  & 31.1 & \textbf{47.3} & 32.3 & 26.2 & 6.5  & 30.1 \\
\midrule
\rowcolor{teal!20} \multicolumn{9}{c}{\textit{Social Sciences \& Humanities}} \\
\midrule
Econ   & 9.5  & 6.0  & 10.0 & 12.6 & 13.5 & 21.1 & 9.8  & \textbf{23.1} \\
Psych  & 6.4  & 8.7  & 15.6 & 18.6 & 20.8 & 23.9 & 7.9  & \textbf{28.5} \\
Phil   & 2.4  & 5.4  & 15.2 & 18.0 & 19.4 & \textbf{21.7} & 7.0  & 18.0 \\
Law    & 10.2 & 19.7 & 30.7 & 35.0 & 35.3 & \textbf{47.6} & 16.4 & 40.2 \\
Christ & 8.9  & 15.0 & 20.0 & 26.5 & 21.0 & \textbf{30.9} & 13.0 & 28.9 \\
Islam  & 12.0 & 10.7 & 25.8 & \textbf{32.0} & 24.3 & 28.9 & 6.5  & 22.6 \\
\midrule
\rowcolor{teal!20} \multicolumn{9}{c}{\textit{Applied Domains}} \\
\midrule
Aviat  & 9.6  & 15.4 & 16.2 & 17.0 & 24.3 & 24.1 & 9.2  & \textbf{29.0} \\
Game   & 17.5 & 19.1 & 41.6 & 43.9 & \textbf{45.6} & 43.1 & 21.4 & 45.2 \\
GIS    & 13.8 & 13.1 & 15.5 & 15.6 & 20.3 & 25.8 & 16.5 & \textbf{34.6} \\
PM     & 8.6  & 8.9  & 21.9 & \textbf{33.2} & 20.5 & 27.6 & 12.4 & 27.1 \\
Sustain& 10.1 & 9.0  & 16.7 & 25.6 & 24.3 & 24.7 & 11.5 & \textbf{28.3} \\
Travel & 10.1 & 16.1 & 23.9 & 30.8 & 26.6 & 36.7 & 13.1 & \textbf{45.7} \\
Quant  & 8.1  & 2.1  & 12.4 & 15.3 & 11.6 & 16.2 & 5.8  & \textbf{23.6} \\
\midrule
\textbf{Average} & 10.0 & 10.4 & 19.5 & 22.0 & 23.0 & 27.6 & 10.8 
& \textbf{29.7} \\
\bottomrule
\end{tabular}%
}
\caption{Per-domain nDCG@10 on MM-BRIGHT (multimodal-to-text track) across 
all 29 domains, grouped by thematic category. BRIDGE achieves the highest 
score in 22 out of 29 domains. \textbf{Bold} denotes the best score per row.}
\label{tab:mmbright_results}
\end{table*}

\subsection{Domain-Level Analysis.}
Table~\ref{tab:mmbright_results} reveals consistent patterns across the 
four domain groups. BRIDGE achieves the best score in \textbf{22 out of 
29 domains}, with the strongest absolute performance in Ubuntu (50.4), 
Travel (45.7), Gaming (45.2), Biology (41.1), and Law (40.2). The largest 
gains over Nomic-Vision arise in domains where screenshots and structured 
visuals carry dense domain-specific information: Ubuntu (+16.1), Biology 
(+14.2), Cryptocurrency (+9.0), and Travel (+9.0). Conversely, BRIDGE underperforms Nomic-Vision in domains where visual content is 
decorative relative to the text question --- Law (40.2 vs.\ 47.6), 
Mathematics (31.9 vs.\ 34.0), and Apple Support (21.3 vs.\ 28.7) --- 
where FORGE gains little signal beyond the text query alone. Quantum 
Computing (8.9) and Robotics (19.0) remain low across all models.

\subsection{Effect of FORGE Backbone.}
\begin{figure}[h]
    \centering
    \includegraphics[width=\linewidth]{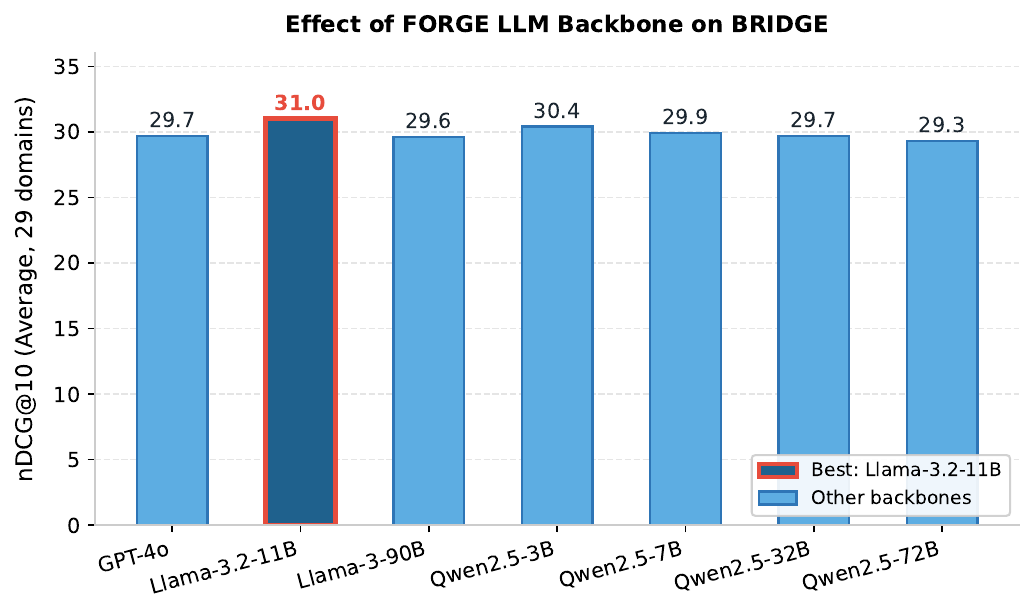}
    \caption{Effect of FORGE LLM backbone on BRIDGE performance (nDCG@10 
    averaged over 29 MM-BRIGHT domains). Llama-3.2-11B achieves the 
    strongest result (31.0), outperforming both larger models (Qwen2.5-72B: 
    29.3) and the GPT-4o zero-shot baseline (29.6). Our default FORGE 
    backbone (Qwen2.5-7B, \textbf{29.7}) offers the best 
    performance-to-cost tradeoff.}
    \label{fig:forge_backbone}
\end{figure}

Figure~\ref{fig:forge_backbone} compares seven FORGE backbones ranging 
from 3B to 72B parameters. \textbf{Llama-3.2-11B achieves the best overall 
performance} (31.0), outperforming both smaller models (Qwen2.5-3B: 30.4) 
and much larger ones (Qwen2.5-72B: 29.3). We attribute this non-monotonic 
scaling to instruction-following quality rather than raw capacity: 
Llama-3.2-11B is optimized for concise, structured generation, which 
aligns naturally with FORGE's task of producing compact search strings, 
whereas larger Qwen2.5 models tend toward verbose outputs that reintroduce 
the lexical noise FORGE is trained to eliminate. GPT-4o used directly as 
a zero-shot rewriter achieves 29.6 --- slightly below our RL-trained 
Qwen2.5-7B (29.7) --- confirming that task-specific RL training on 
retrieval rewards yields competitive performance even against frontier 
models with far more parameters. We adopt \textbf{Qwen2.5-7B} as our 
default FORGE backbone throughout all experiments, as it achieves strong 
performance at manageable inference cost.

\subsection{Ablation: BRIDGE Components.}
\label{sec:ablation}
Table~\ref{tab:ablation} isolates each component's contribution by 
progressively building the full BRIDGE pipeline. Starting from LENS alone 
with raw text queries (25.4), adding the GPT-4o image caption yields a 
+2.4 point gain (27.8), confirming that visual grounding via captioning 
provides useful signal even without alignment. Replacing the raw caption 
concatenation with a supervised FORGE rewriter adds a further +0.7 points 
(28.5), and switching to the full RL-trained FORGE delivers an additional 
+1.2 points to reach \textbf{29.7}. Each component contributes 
incrementally, with the RL training step providing the final and decisive 
improvement over supervised rewriting --- demonstrating that optimizing 
directly for retrieval quality, rather than imitating reference rewrites, 
is essential.

\begin{table}[h]
\centering
\caption{Component ablation on MM-BRIGHT. Each row adds one component 
on top of the previous. Default BRIDGE configuration in \textbf{bold}.}
\label{tab:ablation}
\begin{tabular}{lcc}
\toprule
\textbf{Configuration} & \textbf{nDCG@10} \\
\midrule
LENS only (raw $q_t$, no image)          & 25.4 \\
LENS + Caption (raw $q_t + \delta(q_v)$) & 27.8 \\
LENS + FORGE-SFT (supervised rewriter)   & 28.5 \\
\textbf{LENS + FORGE-RL (full BRIDGE)}   & \textbf{29.7} \\
\bottomrule
\end{tabular}
\end{table}

\subsection{FORGE as a Plug-and-Play Aligner.}
\label{sec:plugin}
Table~\ref{tab:plugin} applies FORGE on top of four different base 
retrievers to test whether the alignment benefit is retriever-agnostic. 
FORGE consistently improves every retriever: BM25 gains +7.3 points 
(8.5 $\to$ 15.8), Nomic-Vision gains +5.7 points (27.6 $\to$ 33.3), 
GME-7B gains +6.3 points (22.0 $\to$ 28.3), and LENS gains +4.3 points 
(25.4 $\to$ 29.7). The gains are largest for weaker base retrievers, 
where the better-formed query compensates for a larger initial retrieval 
gap. Critically, FORGE + Nomic-Vision (33.3) outperforms BRIDGE (29.7) and 
even the best text-only retriever (32.2). However, this combination 
requires a multimodal encoder at inference time, inheriting the 
deployment cost and modality dependency that BRIDGE is explicitly 
designed to avoid. BRIDGE remains the preferred configuration when 
lightweight, encoder-free inference is required; FORGE + Nomic-Vision 
represents the accuracy-optimized variant when inference-time vision 
capacity is available.

\begin{table}[t]
\centering
\caption{FORGE applied to different base retrievers on MM-BRIGHT. 
$\Delta$ = absolute nDCG@10 gain from FORGE alignment.}
\label{tab:plugin}
\begin{tabular}{lccc}
\toprule
\textbf{Base Retriever} & \textbf{Base} & \textbf{+FORGE} & 
$\boldsymbol{\Delta}$ \\
\midrule
BM25                & 8.5  & 15.8 & +7.3 \\
GME-Qwen2-VL-7B     & 22.0 & 28.3 & +6.3 \\
Nomic-Vision        & 27.6 & 33.3 & +5.7 \\
LENS (ours)         & 25.4 & \textbf{29.7} & +4.3 \\
\bottomrule
\end{tabular}
\end{table}

\subsection{FORGE vs.\ Other Query Rewriting Methods.}
\label{sec:rewrite_compare}
Table~\ref{tab:rewrite} compares FORGE against established query rewriting 
baselines, all paired with LENS. For a fair comparison, HyDE and Query2Doc 
are provided with the full concatenated input $x = [q_t \| \delta(q_v)]$, 
identical to FORGE's input. Despite this, HyDE (11.2) and Query2Doc (14.5) 
both underperform even the raw text query (25.4): both methods generate 
hypothetical documents or verbose pseudo-expansions that amplify retrieval 
noise rather than distilling intent, a failure mode that worsens in the 
multimodal setting where the input is already long and entangled. 
FORGE-RL (29.7) substantially outperforms all alternatives, confirming 
that explicit RL training on retrieval rewards is necessary to learn 
effective multimodal query distillation.

\begin{table}[t]
\centering
\caption{Comparison of query rewriting strategies on MM-BRIGHT, all using 
LENS as the base retriever. FORGE-RL denotes our full RL-trained aligner.}
\label{tab:rewrite}
\begin{tabular}{lc}
\toprule
\textbf{Query Method} & \textbf{nDCG@10} \\
\midrule
Raw $q_t$ (no rewrite)          & 25.4 \\
HyDE~\cite{gao2022hyde}         & 11.2 \\
Query2Doc~\cite{wang2023query2doc} & 14.5 \\
\textbf{FORGE-RL (ours)}        & \textbf{29.7} \\
\bottomrule
\end{tabular}
\end{table}

The comparison with GPT-4o zero-shot rewriting (29.6) is particularly 
informative: our RL-trained Qwen2.5-7B FORGE model achieves competitive 
performance despite having 10$\times$ fewer parameters and no access to 
GPT-4o's pretraining data, confirming that RL optimization on task-specific 
retrieval rewards is a more efficient training signal than scale alone.

%% file: sec/6_Conclusion.tex
\section{Conclusion}

We presented \textbf{BRIDGE}, a two-component system that reframes 
multimodal-to-text retrieval as a query representation problem rather 
than a model capacity problem. The core insight is simple: multimodal 
encoders fail not because they lack visual understanding, but because 
the queries they receive entangle visual content, conversational noise, 
and retrieval intent in ways that degrade embedding similarity. BRIDGE 
addresses this at the source through \textbf{FORGE}, an RL-trained query 
alignment model that distills noisy multimodal queries into compact, 
retrieval-optimized search strings, and \textbf{LENS}, a reasoning-enhanced 
dense retriever fine-tuned to handle the intent-rich queries FORGE produces. 
Together, BRIDGE requires no multimodal encoder at inference time and 
operates entirely in the text modality. Evaluated on MM-BRIGHT across 29 domains, BRIDGE achieves \textbf{29.7} 
nDCG@10, surpassing all multimodal encoder baselines including the best 
VLM-based retriever (Nomic-Vision: 27.6). Backbone ablations confirm that 
RL-trained alignment at 7B scale is competitive with GPT-4o zero-shot 
rewriting, and that task-specific training matters more than raw model 
scale. BRIDGE reduces the gap between multimodal and text-only retrieval by 46\% 
relative to the strongest prior multimodal system, demonstrating that the 
remaining distance to the text-only ceiling is a query representation 
problem, not a vision encoding problem. We hope this reframing inspires 
future work to close the gap entirely.

\section*{Acknowledgment}
This work was supported by Innovative Human Resource Development for Local Intellectualization program through the Institute of Information \& Communications Technology Planning \& Evaluation(IITP) grant funded by the Korea government(MSIT) (IITP-2026-RS-2020-II201462, 50\%), and partly supported by the National Research Foundation of Korea (NRF) grant funded by the Korea government (Ministry of Science and ICT) (RS-2023-NR076833)